# Deriving Conceptual Data Models from Domain Ontologies for Bioinformatics

Haya EL-Ghalayini, Mohammed Odeh, and Richard McClatchey
*Centre for Complex Cooperative Systems, Faculty of CEMS, University of the West of England*
*Haya2.Elghalayini@uwe.ac.uk*

Dawn Arnold
*Centre for Research in Plant Science, Faculty of Applied Science, University of the West of England*

**Abstract**

*This paper studies the role that ontologies can play in establishing conceptual data models during the process of information systems development. A mapping algorithm has been proposed and embedded in a special purpose Transformation-Engine to generate a conceptual data model from a given domain ontology. In addition, this paper focuses on applying the proposed approach to a bioinformatics context as the nature of biological data is considered a barrier in representing biological conceptual data models. Both quantitative and qualitative methods have been adopted to critically evaluate this new approach. The results of this evaluation indicate that the quality of the generated conceptual data models can reflect the problem domain entities and the associations between them. The results are encouraging and support the potential role that this approach can play in providing a suitable starting point for conceptual data model development.*

## 1. Introduction

In the past decade, advances in biological experiments have extended from traditional lab work to *in-silico* labs. This has led to an exponential increase in the growth rate of genomics data which is expected to increase in the future. Thus, the need to organize and process these data using computer software technology is an essential need rather than a desirable one.

Bioinformatics is the term that was coined in the mid-1980's to combine both biology and computer sciences and according to the National Institute of Health, Bioinformatics is *"the field of science in which biology, computer science, and information technology merge to form a single discipline. The ultimate goal of the field is to enable the discovery of new biological insights as well as to create a global perspective from which unifying principles in biology can be discerned."* [1].

An effective bioinformatics information system requires the development of a conceptual data model that captures the clear semantics of a certain problem domain for both semantic data modellers and users of the underlying system. Conceptual data models describe the domain problem in semantic terms, using an abstract yet formalized language [2]. These models serve the context of information system (IS) development in communicating between developers and users, helping analysts to understand a domain, providing input to the design process, and documenting the requirements for future reference [2].

However, the difficulties of representing and managing biological data models are specific to the nature of biological data compared to traditional business data. For instance, biological data have complicated structures, high interrelationships to each other, and are of heterogeneous sources [3]. For example, a chromosome can be described in terms of its components: gene, intron, exon, and promoter; sequence: DNA, mRNA, or protein sequence; or structure: centromeres, telomeres and origins of replication. In this situation, it becomes difficult to integrate current legacy systems to infer related data from different data sources since each data source has its own syntactic formats as well as semantic interpretation.

The high complexity in understanding the biological data leads researchers to propose different approaches in simplifying the process of modelling biological data. A number of researchers such as Bornberg-Bauer and Paton [4] among others propose a set of conceptual data models for a range of important emerging information resources in bioinformatics which are represented by the most widely used conceptual modelling techniques such as entity-relationship (ER) or UML. Hence, the previous mentioned works emphasize the important role that conceptual data models can play in resolving difficulties in describing and understanding the structure of biological data.

As ontologies support *"a shared and common understanding of a domain that can be communicated between people and across application systems"* [5], their importance is also recognized within the bioinformatics community [6], for example, to provide applications with domain knowledge and to thereby facilitate the interoperability between heterogeneous data sources, to describe a database schema or to annotate biological data.

This paper reports on the role that ontologies can play in developing the respective conceptual data models during the process of information systems development. In particular this has led to devise a mapping algorithm in a form of a so-called ***Transformation-Engine (TE)*** to generate a possible conceptual data model from a given domain ontology. To evaluate the extent of the accuracy of the generated conceptual data model (GCDM), we chose to apply this transformation using a certain ontology representation language, the Web Ontology

Language-OWL [7], and an application domain, the TAMBIS bioinformatics ontology-TaO, which is specialized in molecular biology concepts [8]. We chose to study ontologies represented by OWL, since it is the most recent web ontology language released by the World Wide Web Consortium and since its formal semantics are based on description logics (DL).

The rest of the paper is organized as follows: Section 2 gives a brief introduction on the approach of reverse engineering domain ontologies to conceptual data models. This approach is demonstrated by a real-life case study related to the bioinformatics ontology in Section 3. Section 4 discusses the process of evaluating the quality of the GCDMs that are derived from the domain ontologies. Finally, we conclude and suggest future research works in Section 5.

## 2. Reverse engineering domain ontologies to conceptual data models approach

There are some similarities and differences between ontologies and conceptual data models. Both are represented by a modelling grammar with similar constructs, such as classes in ontologies which correspond to entity types in conceptual data models (CDM). Thus, the methodologies of developing both models have common activities [9]. While ontologies and conceptual data models share common features, they do have some differences. Fonseca [9] defines two criteria that differentiate ontologies from conceptual data models; the first is the objectives of modelling and the second is objects to model. Using the first criterion, an ontology focuses on the description of the "invariant features that define the domain of interest", whereas a conceptual data model links the domain invariant features with a set of observations to be defined within an information system. Regarding the second criterion, objects to model, an ontology describes real or factual structures of a domain which enables information integration. Conversely, a conceptual data model object represents a general category of a certain domain linked to its individual events, for example, linking the general category of gene with the size of its DNA sequence. The central question addressed in this research is:

"To what extent can domain ontologies participate in developing conceptual data models?"

Having surveyed the literature, the differences between ontologies and conceptual data models have mainly been explored using descriptive studies. Thus, in order to address the main research question, a two phase approach has been devised to integrate both interpretative and empirical studies. In the first phase, the ontological model provided by Wand and Weber [10], which is known as the Bunge-Wand-Weber ontology (BWW), has been utilised in interpreting the OWL ontology language. We note that ontology language constructs are related to the structural components of the problem domain. Other constructs related to time dependency have not been represented in OWL. The second phase implements a new algorithm (implemented as a *TE* component) that maps a domain ontology expressed in OWL to a GCDM represented as a UML class model [11]. The process of developing the conceptual data model begins by selecting an OWL ontology of the domain of interest. Then, the *TE* applies the mapping rules onto the ontology concepts, thereby generating sub-models that are integrated to construct the proposed conceptual data model [12]. Briefly the *TE* mapping rules are:

**Rule1**. *The atomic concept in OWL (defined by <owl:class>) is mapped into the entity type in the GCDM.*

**Rule2**. *The OWL subsumption relationship construct between atomic concepts are transformed into a generalization/specialization relationship between entity types in the GCDM.*

**Rule3**. *The logical expression concept in OWL is transformed into a generalization/specialization relationship between the entity types in the GCDM. For example, the <owl:intersectionOf> expression is translated to a multiple inheritance relationship between the operands of the logical expression (as super entity types) and the concept being studied as a sub entity type, whereas the <owl:unionOf> expression partitions the concept being studied, i.e. the super entity type into its operands as sub entity types.*

**Rule4**. *The ontology restriction is transformed to a mutual uni-directional relationship (association) if the restriction concept restricts the link between atomic concepts using a property of type <owl:objectProperty>, namely the concept being studied as a source entity type and the concept being restricted as a target entity type with a multiplicity constraint at its end.*

**Rule5**. *The ontology restriction is transformed to an intrinsic property of an entity type being studied with a defined data type if and only if a restriction concept restricts the property of a concept using <owl:dataTypeProperty> property type.*

**Rule6**. *An OWL mutual property <owl:objectProperty> is transformed to a uni-directional relationship between its domain and range entity types.*

**Rule7**. *An OWL intrinsic property <owl:dataTypeProperty>(P) is transformed to an attribute of the domain entity type with an associated data type.*

**Rule8**. *The TE transforms the inverse relation <owl:inverseOf> of a mutual property <owl:objectProperty> to a bi-directional relationship between the domain and range entity types.*

**Rule9**. *The TE transforms the functional property <owl:functionalProperty> to restrict a relationship between the domain and range entity types with an optional unique constraint (0..1) at the target entity type end.*

**Rule10**. *Redundant relationships with different multiplicity constraints are merged into one relationship with a multiplicity constraint resulting from the intersection of different multiplicity constraints-(Refining Rule).*

***Rule11.*** *A relationship with a defined data type as a target entity type is refined as an attribute of the source entity type with the associated data type-(Refining Rule).*

## 3. Applying the TE to a bioinformatics ontology

The TAMBIS ontology contains knowledge about bioinformatics and molecular biology concepts and their relationships. It describes proteins and enzymes, as well as their motifs, secondary and tertiary structure functions and processes [8]. We use the TAMBIS ontology (TaO) to demonstrate our approach. TaO has 393 concepts and 94 properties whereas the GCDM has 392 entity types, 259 relationships, 49 attributes, and 402 generalization/specialization relationships. In this test case, we have selected the concepts that are relevant to proteins in order to generate the protein sequence CDM using the ***TE***. The GCDM has been translated to a set of Java files and reverse-engineered to a class diagram by using a UML graphical tool.

In what follows, we describe the process of applying the ***TE mapping algorithm*** to TaO for developing the *protein* sequence model. The model includes protein information such as their sequence, function, and structure. The main activities of the ***TE*** are: starting the process, ***TE*** rules computation, and end of the process.

### 3.1. Starting the process

Conceptual data models use only specific and narrow parts of knowledge from a given domain ontology, thus they use only subsets of the domain concepts and properties. The objective of this activity is to identify a set of candidate concepts or "seeds" from TaO to find relevant concepts of the protein sequence. The concepts that are selected as seeds concepts are included in a collection set that contains the following concepts: *{DNA, mRNA, Protein, nucleic-acid, protein function, protein structure, organism classification}*.

### 3.2. The TE rules computation

The computation proceeds by applying the embedded rules of the ***TE*** to the selected seeds concepts. In what follows, we explain the transformation of a seed concept, for example, *protein,* using the ***TE rules.*** After submitting *protein* as a seed concept, the ***TE*** obtains all *protein* relevant concepts and properties in addition to all *protein* sub-concepts that will be added to the collection list. *Protein* is described as a defined concept and as a sub-concept of different description concepts.

#### 3.2.1. Protein as a defined concept

In description logic (DL) formalism, a defined concept defines a set of necessary and sufficient axioms to be used by reasoning services to classify any concept or instance which fulfills these axioms as a sub-concept of the defined concept. Hence *Protein* in DL syntax is described as:

$$Protein \equiv macromolecular\text{-}compound$$
$$\exists Polymer\text{-}of.\,amino\text{-}acid$$
$$\forall Polymer\text{-}of.\,amino\text{-}acid$$

Thus, the ***TE*** applies its rules to deconstruct and transform the above description to the following:
- *Protein, Macromolecular,* and *Amino-Acid* are entity types using ***Rule1***.
- *Protein* is a sub entity type of *Macromolecular* using ***Rule2.***
- *Polymer-of* is a relationship between *Protein* as a source entity type and *Amino-Acid* as a target entity type with a multiplicity constraint one-to-many using ***Rule4*** as an existential ($\exists$) restriction-R1.
- *polymer-of* is a relationship between *Protein* as a source entity type and *Amino-Acid* as a target entity type with a multiplicity constraint zero-to-many using ***Rule4*** as a universal ($\forall$) restriction-R2***.***

Consequently, ***Rule10*** is applied to find the intersection of different multiplicity constraints for the same relationship, i.e. (R1 and R2), thereby refining the multiplicity constraint of *polymer-of* relationship to one-to-many at the end of *Amino-Acid* type.

#### 3.2.2. Description concepts of protein

To define the semantics of *Protein*, OWL also permits the use of concept descriptions such as restriction concepts and logical expression concepts. These concepts are known as anonymous concepts (i.e.they have no names) and are considered as super concepts of *Protein*. Next, we present some examples on transforming the description concepts associated with *Protein* concept using the ***TE*** rules.

***Description Concept 1.***
*Protein* $\subseteq$
($\exists translated from.(DNA \cup mRNA)) \cap$  operand-1
($\forall translated\text{-}from.(DNA \cup mRNA))$  operand-2

*Protein* is described as a sub-concept of *operand-1* where it has a filler that is the union ($\cup$) of *DNA* and *mRNA* concepts. The closure restriction on property *translated-from* is used as *operand-2*. The closure restriction acts along the *translated-from* property to say that it can only be filled by the filler *DNA* and *mRNA*. The intersection ($\cap$) concept between ($\forall$) and ($\exists$) restrictions means that *Protein* must have at least one value of *translated-from* property of *DNA* and *mRNA*. Thus the ***TE*** uses ***Rule4*** ($\forall$ *and* $\exists$ *restrictions*), ***Rule6*** (*translated-from* as an object property) and ***Rule10*** to transform this description to:

- *Translated-from* is a relationship between *Protein* as a source entity type and *DNA* and *mRNA* as target types with a multiplicity constraint one-to-many.

***Description Concept 2.***
*Protein* $\subseteq$  (*has-species* $\geq 1) \cap$   operand-1
($\exists has\text{-}species.Species) \cap$   operand-2
($\forall has\text{-}species.Species$)   operand-3

In this case, *Protein* is described as a sub-concept of an intersection expression that links three operands. *Operand-1* is a cardinality restriction ($\geq$) that restricts the number of possible values of *has-species* property to at least one value. *Operand-2* is a (∃) that restricts the existence of at least one value of *has-species* from *Species* concept. *Operand-3* is a universal restriction that restricts the value of *has-species* to be only from *Species*. After deconstructing the expression into its operands, the **TE** uses **Rule4** (≥, ∀ and ∃ restrictions), **Rule6** (*has-species* as an object property) and **Rule 10** to generate:

- *has-species* is a relationship between *Protein* as a source entity type and *Species* as a target entity type with a multiplicity constraint one-to-many.

**Description Concept 3.**
Protein ⊆
has-sequence≥ 1∩                              operand 1
(∃has-sequence.(Protein-sequence∪
        ∀part-of.Protein-sequence))           operand 2

OWL as an expressive language provides descriptions of complex concepts that are difficult to represent in conceptual data models. The complex concept in the above description is the (∩) concept between two operands representing ≥ and ∃ restrictions. However, the filler of the second operand has a filler of type (∪) expression between the atomic concept and ∀ restriction. This means that for each instance *x* of *Protein* there exists at least one value of the *has-sequence* relationship of type *Protein-sequence* or of type ∀ *restriction*. In other words, defining ∀ *restriction* as a type means that for each instance *x* of *Protein* there exists a relationship *sequence-of(x, y)* such that *y* is an instance of ∀*part-of.Protein-sequence*. Consequently, representing ∀*part-of.Protein-sequence* as an entity type adds more details and complexity to the conceptual data model. Then, for the conceptual data model purposes, the capacity of the **TE** to transform high-expressive concepts and nested descriptions is limited in transforming these nested concepts. Therefore, it will be sufficient to deconstruct these concepts into their operands and consider them as additional relations associated with the concept being studied. Thus, the **TE** transforms the expression to:

- *has-sequence* is a relationship between *Protein* as a source entity type and *Protein-sequence* as a target entity type with a one-to-many multiplicity constraint using **Rule3** (∩ expression), **Rule 4** (≥, ∃ *restriction*), **Rule6** *(has-sequence as an object property)* and **Rule10**.
- *Part-of* is a bi-directional relationship between *Protein* as a source entity type and *Protein-sequence* as a target entity type with zero-to-many multiplicity constraint using **Rule3** (∪ expression), **Rule4** *(∀ restriction)*, **Rule6** *(part-of as an object property)*, **Rule8** *(part-of* has an inverse axiom) and **Rule10**.

Transforming *Protein* and all its concept descriptions generates one generalization/specialization relationship, and 30 relationships between *Protein* and different entity types. Applying the refining rules in the **TE** refactors the GCDM. In the above example the refining rules merge the duplicated relationships thereby generating 16 equivalent relationships instead of 30.

### 3.3. The End of the process

After generating the conceptual data model for the protein sequence and all its relevant concepts, the **TE** repeats the computation process for all concepts in the collection set. At the end of each iteration, the **TE** merge algorithm is executed using two conceptual data models. After that, the process ends by transforming all the concepts in the collection set and merges all individual conceptual data models into one model. The final representation of the *Protein sequence* conceptual data model is illustrated in Figure 1.

## 4. Evaluating the TE mapping rules

### 4.1 Evaluation methodology

The evaluation process of this research aims to measure the quality of the GCDMs that are developed from reverse engineering domain ontologies. The process of evaluation embodies two components, both quantitative and qualitative evaluation methods. We believe that a quantitative evaluation is essential to evaluate the performance of the **TE** mapping algorithm that constitutes the mapping rules. Also, t qualitative evaluation method is proposed to evaluate the adequacy of the GCDM as a conceptualization of a certain problem domain. While quantitative evaluation can be performed using the well established recall and precision, the qualitative evaluation is more subtle and there are no standard methods for performing such evaluations. A common approach is to involve a domain specialist in assessing the quality of the GCDM elements which determines the fitness of the GCDM in some particular task. Thus, the observations of the domain specialist on the GCDM being studying are applied to the GCDM to develop a Gold Model (GM). Consequently, the GM is considered as a valid and complete model for representing a specific real-world phenomenon. We have defined two metrics for evaluating the quality of the GCDM.

- Recall reflects the validity of the GCDM produced by the **TE**. The correct and relevant information in the GCDM is compared with the GM using the following formula:

    Recall = Number(correct) / Number(GM)

where Number(correct) refers to the number of correct and relevant elements in the GCDM, and Number(GM) is the number of information elements in the GM.

- Precision reflects the completeness of the GCDM produced by the **TE**, i.e. how much of the information produced by the **TE** was correct compared to the information produced in the GM. The following formula is used to calculate Precision:

    Precision = Number(correct) / Number(GCDM)

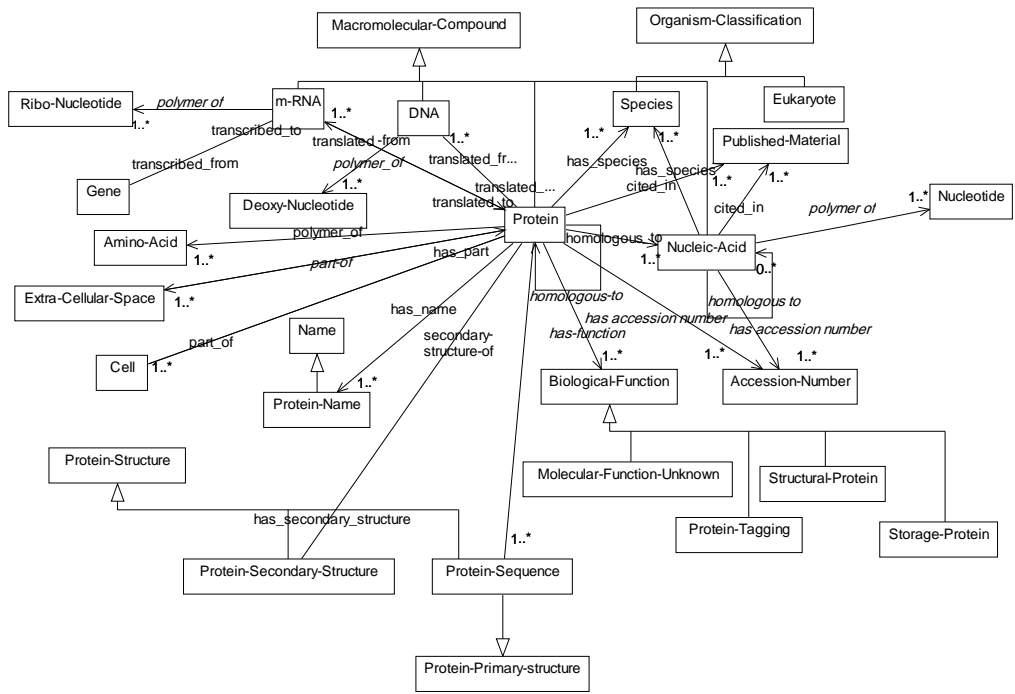

**Figure 1.** *Protein-sequence* conceptual data model

where Number(correct) is as above, and Number(GCDM) is the number of information elements in the GCDM.

To verify the fitness of the GCDM in some particular task, two GCDMs were submitted for evaluation to a domain specialist, a bioinformatics specialist from the University of the West of England. The specialist is familiar with using various bioinformatics data sources and has been presented with an introduction to the basic elements of the conceptual data models. The protein sequence (the protein sequence model described above) and gene-chromosome models are selected to be reviewed and assessed by the expert. The evaluation process involved the correctness of the existing GCDM entities, including their attributes, and relationships with other entities. Therefore, additions and deletions for some elements of the GCDM have been placed to obtain the GM.

### 4.2 Evaluation results

The average Recall of at least 90% and Precision of 74% -80% for both the entities and the relations in the GCDM reflect the high degree of adequacy of the GCDM in identifying the substantial problem domain entities along with their relationships. In other words, the results indicate that the semantic quality of the GCDM from a given domain ontology can be very reflective of the problem domain entities and the relations between them. Therefore, the semantics of the GCDMs conform strongly to the domain consensual knowledge with the important distinction that these models are not developed for specific applications.

The following conclusions can be drawn from the observations of the domain expert:

• The specialist concluded that some knowledge specified by a given domain ontology is insufficiently precise for transformation into conceptual data model elements. This is because in some cases the domain ontology is oriented towards a specific task to be carried out in the domain rather than task independent domain knowledge. For example, TaO states that *DNA* is translated into *Protein*, but this is not precise as *mRNA* is translated to *Protein*. Therefore, these concepts must be updated in the GM to represent a clear and unambiguous representation of the underlying domain problem.

• As predicted, unexpected elements are presented in the GCDM that are not considered as substantial entities such as *Macromolecule-Part*, *Genome-Part*, *Gene-part*, or *DNA-Part.* This is because the OWL language as well as other DL languages uses 'open world reasoning' that supports defined concepts which provide sufficient and necessary conditions for classifying other concepts as sub-concepts once they fulfill the super-concept conditions. The aim of defining these concepts is to support representing ontology concepts in a taxonomical structure; consequently, new information is derived. Therefore, the ambiguity in translating these concepts using the ***TE*** does exist as they can be used originally either for describing the main properties of substantial concepts, i.e. defined concepts can be represented as substantial entities in the GCDM or clarifying the meaning of a group of substantial concepts, i.e. the defined concept cannot be represented as substantial entities in the GCDM). For example, *DNA-Part* is defined in TaO as:

*DNA-Part* ≡ *Macromolecule-Part*
$\subseteq$ ((*Part-of.DNA*) ∩ (*Part-of.DNA*))

where its purpose is to clarify the meaning of *DNA-Part* by classifying any *Macromolecule-Part* type which is *part-of DNA* as a sub-concept of *DNA-Part* type using a reasoning service. Conceptually, *DNA-Part* is not a substantial entity and cannot be represented as an entity type in the GCDM; therefore, it has to be eliminated in the GCDM since it does not add valuable meaning to the structure of the data in the problem domain.

• Overall, the domain expert accepted the two models. Furthermore, the average amount of human intervention for additional refinements (i.e. for deriving the GMs) that were carried out by some form of minimal communication with domain specialist compared to developing these models from scratch are 20% for the entities and 14% for the relationships.

However, there still remains the question that needs to be answered in relation to the misrepresented elements in the GCDMs, even if they are not statistically significant. This observation stems from the fact that the OWL class construct is overloaded to represent dynamic and static real world characteristics, i.e. the same construct is used to represent a domain concept, event, process, or transformation. To overcome this problem, we suggest extending the definition of the class construct to incorporate a meta-concept element to distinguish between these different concepts. Therefore the 'static' meta-concept represents domain concepts which identify and support the identity property of an entity type, i.e. substantial entities, whereas the 'dynamic' meta-concept represents an event or transformation concept that captures the behavior of a given real world phenomenon.

## 5. Conclusions & future work

This paper has presented a new approach that has been developed to automate the derivation of the GCDMs from domain ontologies. And, to evaluate the extent of the accuracy of the GCDM, we chose to apply the *TE* to OWL as the recent web ontology language and the TAMBIS bioinformatics ontology. The results of the evaluation process prove the effectiveness of the approach to alleviate the complexity in understating and identifying the biological entities and the associations with other entities of a particular problem domain. Indeed, the semantics of the GCDMs conform strongly to the domain consensual knowledge with the important distinction that these models are not developed for specific applications.

Finally, this research has been focused on using a single domain ontology to generate some highly accurate respective conceptual data models. But, as different ontologies within the same domain capture common as well as different knowledge for different purposes, this has influenced this research to further investigate the need to develop an approach to integrate related domain ontologies in order to obtain a unified domain ontology [3].